\documentstyle[twoside,epsf]{article}

\catcode`\@=11
\long\def\@makefntext#1{
\protect\noindent \hbox to 3.2pt {\hskip-.9pt  
$^{{\eightrm\@thefnmark}}$\hfil}#1\hfill}               

\def\@makefnmark{\hbox to 0pt{$^{\@thefnmark}$\hss}}    
        
\def\ps@myheadings{\let\@mkboth\@gobbletwo
\def\@oddhead{\hbox{}
\rightmark\hfil\eightrm\thepage}   
\def\@oddfoot{}\def\@evenhead{\eightrm\thepage\hfil
\leftmark\hbox{}}\def\@evenfoot{}
\def\sectionmark##1{}\def\subsectionmark##1{}}

\oddsidemargin=\evensidemargin
\newcounter{sectionc}\newcounter{subsectionc}\newcounter{subsubsectionc}
\renewcommand{\section}[1] {\vspace{12pt}\addtocounter{sectionc}{1} 
\setcounter{subsectionc}{0}\setcounter{subsubsectionc}{0}\noindent 
        {\tenbf\thesectionc. #1}\par\vspace{5pt}}
\renewcommand{\subsection}[1] {\vspace{12pt}\addtocounter{subsectionc}{1} 
\setcounter{subsubsectionc}{0}\noindent 
{\bf\thesectionc.\thesubsectionc. {\kern1pt \bfit #1}}\par\vspace{5pt}}
\renewcommand{\subsubsection}[1] {\vspace{12pt}\addtocounter{subsubsectionc}{1}
        \noindent{\tenrm\thesectionc.\thesubsectionc.\thesubsubsectionc.
        {\kern1pt \tenit #1}}\par\vspace{5pt}}
\newcommand{\nonumsection}[1] {\vspace{12pt}\noindent{\tenbf #1}
        \par\vspace{5pt}}

\newcounter{appendixc}
\newcounter{subappendixc}[appendixc]
\newcounter{subsubappendixc}[subappendixc]
\renewcommand{\thesubappendixc}{\Alph{appendixc}.\arabic{subappendixc}}
\renewcommand{\thesubsubappendixc}
        {\Alph{appendixc}.\arabic{subappendixc}.\arabic{subsubappendixc}}

\renewcommand{\appendix}[1] {\vspace{12pt}
        \refstepcounter{appendixc}
        \setcounter{figure}{0}
        \setcounter{table}{0}
        \setcounter{lemma}{0}
        \setcounter{theorem}{0}
        \setcounter{corollary}{0}
        \setcounter{definition}{0}
        \setcounter{equation}{0}
        \renewcommand{\thefigure}{\Alph{appendixc}.\arabic{figure}}
        \renewcommand{\thetable}{\Alph{appendixc}.\arabic{table}}
        \renewcommand{\theappendixc}{\Alph{appendixc}}
        \renewcommand{\thelemma}{\Alph{appendixc}.\arabic{lemma}}
        \renewcommand{\thetheorem}{\Alph{appendixc}.\arabic{theorem}}
        \renewcommand{\thedefinition}{\Alph{appendixc}.\arabic{definition}}
        \renewcommand{\thecorollary}{\Alph{appendixc}.\arabic{corollary}}
        \renewcommand{\theequation}{\Alph{appendixc}.\arabic{equation}}
        \noindent{\tenbf Appendix \theappendixc #1}\par\vspace{5pt}}
\newcommand{\subappendix}[1] {\vspace{12pt}
        \refstepcounter{subappendixc}
        \noindent{\bf Appendix \thesubappendixc. {\kern1pt \bfit #1}}
        \par\vspace{5pt}}
\newcommand{\subsubappendix}[1] {\vspace{12pt}
        \refstepcounter{subsubappendixc}
        \noindent{\rm Appendix \thesubsubappendixc. {\kern1pt \tenit #1}}
        \par\vspace{5pt}}

\newcommand{\textlineskip}{\baselineskip=13pt}
\newcommand{\smalllineskip}{\baselineskip=10pt}

\newcommand{\copyrightheading}[1]
        {\vspace*{-2.5cm}\smalllineskip{\flushleft
        {\footnotesize Quantum Information and Computation, Vol.~3, No.~2 (2003) 121--138 #1}\\
        {\footnotesize \copyright\kern2pt Rinton Press}\\
         }}

\newcommand{\publisher}[2]{{\begin{center}\footnotesize\smalllineskip 
        Received #1\\
        Revised #2
        \end{center}
        }}

\def\abstracts#1#2#3{{
        \centering{\begin{minipage}{4.5in}\footnotesize\baselineskip=10pt
        \parindent=0pt #1\par 
        \parindent=15pt #2\par
        \parindent=15pt #3
        \end{minipage}}\par}} 

\def\keywords#1{{
        \centering{\begin{minipage}{4.5in}\footnotesize\baselineskip=10pt
        {\footnotesize\it Keywords}\/: #1
         \end{minipage}}\par}}
\def\communicate#1{{
        \centering{\begin{minipage}{4.5in}\footnotesize\baselineskip=10pt
        {\footnotesize\it Communicated by}\/: #1
         \end{minipage}}\par}}

\renewenvironment{thebibliography}[1]
        {\frenchspacing
         \ninerm\baselineskip=11pt
         \begin{list}{\arabic{enumi}.}
        {\usecounter{enumi}\setlength{\parsep}{0pt}     
         \setlength{\leftmargin 12.7pt}{\rightmargin 0pt}
         \setlength{\itemsep}{0pt} \settowidth
        {\labelwidth}{#1.}\sloppy}}{\end{list}}

\newcounter{itemlistc}
\newcounter{romanlistc}
\newcounter{alphlistc}
\newcounter{arabiclistc}

\newcommand{\fcaption}[1]{
        \refstepcounter{figure}
        \setbox\@tempboxa = \hbox{\footnotesize Fig.~\thefigure. #1}
        \ifdim \wd\@tempboxa > 5in
           {\begin{center}
        \parbox{5in}{\footnotesize\smalllineskip Fig.~\thefigure. #1}
            \end{center}}
        \else
             {\begin{center}
             {\footnotesize Fig.~\thefigure. #1}
              \end{center}}
        \fi}

\newcommand{\tcaption}[1]{
        \refstepcounter{table}
        \setbox\@tempboxa = \hbox{\footnotesize Table~\thetable. #1}
        \ifdim \wd\@tempboxa > 5in
           {\begin{center}
        \parbox{5in}{\footnotesize\smalllineskip Table~\thetable. #1}
            \end{center}}
        \else
             {\begin{center}
             {\footnotesize Table~\thetable. #1}
              \end{center}}
        \fi}

\def\pmb#1{\setbox0=\hbox{#1}
        \kern-.025em\copy0\kern-\wd0
        \kern.05em\copy0\kern-\wd0
        \kern-.025em\raise.0433em\box0}

\def\fnt#1#2{\footnotetext{\kern-.3em
        {$^{\mbox{\scriptsize #1}}$}{#2}}}

\def\fpage#1{\begingroup
\voffset=.3in
\thispagestyle{empty}\begin{table}[b]\centerline{\footnotesize #1}
        \end{table}\endgroup}

\def\runninghead#1#2{\pagestyle{myheadings}
\markboth{{\protect\footnotesize\it{\quad #1}}\hfill}
{\hfill{\protect\footnotesize\it{#2\quad}}}}
\font\tenrm=cmr10
\font\tenit=cmti10 
\font\tenbf=cmbx10
\font\bfit=cmbxti10 at 10pt
\font\ninerm=cmr9

\font\eightrm=cmr8

\def\FigName{figure}%
\newbox\captionbox
\long\def\@makecaption#1#2{%
  \ifx\FigName\@captype
    \vskip\abovecaptionskip
    \setbox\tempbox\hbox{{\figurecaptionfont #1\hskip1em #2}}
        \ifdim\wd\tempbox< 28pc
        \centerline{\box\tempbox}
        \else
        {\figurecaptionfont #1\hskip1em #2\par}
\fi\else
        \setbox\tempbox\hbox{{\tablecaptionfont #1\hskip1em #2}}
        \ifdim\wd\tempbox< 28pc 
        \centerline{\box\tempbox}
        \else
        {\tablecaptionfont #1\hskip1em #2\par}%
        \fi   
 \vskip\belowcaptionskip
 \fi}
\InputIfFileExists{psfig.sty}
{\typeout{^^Jpsfig.sty inputed...ok}}{\typeout{^^JWarning: psfig.sty could be be found.^^J}}
\InputIfFileExists{epsfsafe.tex}
{\typeout{^^Jepsfsafe.tex inputed...ok}}
                        {\typeout{^^JWarning: epsfsafe.tex could not be found.^^J}}
\InputIfFileExists{epsfig.sty}
{\typeout{^^Jepsfig.sty inputed...ok}}{\typeout{^^JWarning: epsfig.sty could not be found.^^J}}
\InputIfFileExists{epsf.sty}
{\typeout{^^Jepsf.sty inputed...ok}}{\typeout{^^JWarning: epsf.sty could not be found.^^J}}%
\def\fps@figure{tbp}
\def\ftype@figure{1}
\def\ext@figure{lof}
\def\fnum@figure{Fig.\ \thefigure}
\def\qed{\hbox{${\vcenter{\vbox{                  
   \hrule height 0.4pt\hbox{\vrule width 0.4pt height 6pt
   \kern5pt\vrule width 0.4pt}\hrule height 0.4pt}}}$}}

\begin{document}
\setlength{\textheight}{8.0truein}    

\runninghead{An analysis of reading out the state of a charge quantum bit} 
            {H.-S. Goan}

\normalsize\textlineskip
\thispagestyle{empty}
\setcounter{page}{121}

\copyrightheading{}      

\vspace*{0.88truein}

\fpage{1}
\centerline{\bf
AN ANALYSIS OF READING OUT THE STATE OF}
\centerline{\bf A CHARGE QUANTUM BIT}
\vspace*{0.035truein}
\vspace*{0.37truein}
\centerline{\footnotesize 
HSI-SHENG GOAN\footnote{Mailing address: Center for Quantum Computer
Technology, C/- Department of Physics, University of Queensland,
Brisbane 4072, Australia; E-mail: goan@physics.uq.edu.au}}
\vspace*{0.015truein}
\centerline{\footnotesize\it Center for Quantum Computer Technology, 
University of New South Wales
}
\baselineskip=10pt
\centerline{\footnotesize\it Sydney, NSW 2052 Australia}
\publisher{September 9, 2002}{November 13, 2002}

\vspace*{0.21truein}
\abstracts{
We provide a unified picture for the master equation approach 
and the quantum trajectory approach to a measurement problem
of a two-state quantum system (a qubit),
an electron coherently tunneling between two coupled quantum dots
(CQD's) measured by a low transparency point contact (PC) detector.
We show that the master equation of ``partially'' reduced density 
matrix can be derived from the 
quantum trajectory equation (stochastic master equation)
by simply taking a ``partial'' average 
over the all possible outcomes of the measurement.
If a full ensemble average is taken, 
the traditional (unconditional) master equation of
reduced density matrix is then obtained. 
This unified picture, in terms of averaging over (tracing out) 
different amount of detection records (detector states), 
for these seemingly
different approaches reported in the literature
is particularly easy to understand using
our formalism. To further demonstrate this connection, we analyze an
important ensemble quantity for an initial qubit state readout experiment,
$P(N,t)$, the probability distribution of
finding $N$ electrons that have 
tunneled through the PC barrier in time $t$. 
The simulation results of $P(N,t)$ using 10000 quantum trajectories
and corresponding measurement records are, as expected, in very good
agreement with those obtained from the Fourier analysis of 
the ``partially'' reduced density matrix.
However, the quantum trajectory approach provides more information and
more physical insights into the ensemble and time
averaged quantity $P(N,t)$.   
Each quantum trajectory resembles a single history of the qubit state
in a single run of the continuous measurement experiment. 
We finally discuss, in this approach, the possibility
of reading out the state of the qubit system 
in a single-shot experiment.}{}{}

\vspace*{10pt}
\vspace*{3pt}
\communicate{D Wineland \& P Delsing}

\vspace*{1pt}\textlineskip      
\section{Introduction}          
\vspace*{-0.5pt}
\noindent
In condensed matter physics measurements are usually performed in two ways. 
Either a measurement is performed on an ensemble of quantum
systems. Or for individual quantum system, only one measurement is
made in each experimental run, but the same experiments are repeated many
times. In the above two cases, only ensemble average properties 
are studied. Hence, the traditional quantum open system approach, namely the
(unconditional) master equation of the reduced density matrix obtained
by tracing out the environmental (measurement apparatus) degrees of
freedom, is sufficient to describe the ensemble average time
evolution of the system of interest. 
However, when a sequence of measurements in an
experiment is made upon a single quantum system, the system state conditioned
upon the measurement result is important when its
subsequent time evolution is concerned. So if one wants to map out the
system state evolution conditioned on the continuous in time
measurement results for a single quantum system (i.e. quantum
trajectory of the measured system), the conditional, 
stochastic Schr\"odinger (master)
equation or quantum trajectory theory \cite{QT}
should be employed. This is
indeed sometimes the case for the readout process of a single quantum
bit (qubit) in various proposed solid-state quantum computer
architectures \cite{Kane98,Loss98,Schoen97,Averin98,Tanamoto00}.

The theory of quantum trajectories or stochastic Schr\"odinger
equations \cite{QT} has been developed
in last ten years 
mainly in the quantum optics
community to describe open
quantum system subject to continuous quantum measurements. 
But it was introduced to the context of
solid-state mesoscopics only recently
\cite{Korotkov99,Korotkov01b,Goan01a,Goan01b}
to discuss a two-state quantum system (a charge qubit), 
an electron coherently tunneling between two coupled quantum dots (CQD's),
interacting with an environment (a detector), a low transparency point
contact (PC) or tunnel junction \cite{Gurvitz97}.
One of the main purposes of the paper is to provide a unified picture for the 
quantum trajectory approach and the 
master equation approach of reduced and ``partially'' 
reduced density matrix.
Here we refer to the master equation approach of the ``partially'' 
reduced density matrix as the  
approach recently developed in Refs.\ 
\cite{Gurvitz97,Shnirman98,Gurvitz98,Makhlin00,Makhlin01}, 
mainly for the purpose of 
reading out an initial qubit state. 
In that approach, one takes a trace over
environmental (detector) microscopic degrees of the freedom but 
keeps track of the number of electrons, $N(t)$, that have tunneled
through the PC barrier during time $t$ in the reduced density matrix.
In Refs.\cite{Korotkov99,Korotkov01b,Goan01a,Goan01b},
the difference and connection between 
the (conditional) quantum trajectory approach and the (unconditional) master
equation approach of reduced density matrix were emphasized. But, to our
knowledge, no direct connection between the quantum trajectory approach and
the master equation 
approach of the "partially" reduced density matrix has been formally
established and reported in the literature so far. 
One of the reasons may be that the methods
used to derive the stochastic quantum trajectory 
equations and master equation of the
``partially'' reduced density matrix are so different.
The master equation (rate equations) of the ``partially'' 
reduced density matrix for the
CQD qubit system measured by a PC detector 
was derived in Ref.\ \cite{Gurvitz97} 
from the so-called many-body Schr\"odinger
equation. While it was derived in 
Refs.\ \cite{Shnirman98,Makhlin00} and \cite{Makhlin01}, 
by means of the 
diagrammatic technique
in the Keldysh forward and backward in time contour,
for a Cooper-pair-box charge qubit coupled
capacitively to a single-electron transistor detector.      
On the other hand, the prescriptions for the quantum
trajectories (or conditional, stochastic system state evolution)
in Refs.\ \cite{Korotkov99} and \cite{Korotkov01b} 
were based on the Bayesian
formalism, while they
were derived in Refs.\ \cite{Goan01a} and 
\cite{Goan01b} starting
from unconditional master equation.

However, in this paper
we provide a direct connection between the quantum
trajectory approach and the ``partially'' reduced density matrix approach. 
We show that the master equation of the ``partially'' reduced 
density  can be obtained for the CQD/PC model by taking a
``partial'' average on the conditional, stochastic 
master equation of the CQD qubit system 
density matrix over the possible outcomes of the
measurements of the PC detectors.
The procedure to achieve this connection is particularly easy to
understand using our formalism.
In fact,
the quantum trajectories provide us with the most (all)
information as far as the system evolution is concerned. When more and
more information of the detector states (detection records) 
is traced (averaged) 
out from the quantum trajectories, we obtain first
the ``partially'' reduced 
density matrix and then the reduced density matrix.
This provides a unified picture for these seemingly different approaches
reported in the literature.

To further demonstrate the connection between the quantum trajectory
approach and the ``partially'' reduced density matrix approach, we
consider an experiment of reading out an initial CQD qubit state.
We analyze an important ensemble quantity in the readout
experiment, $P(N,t)$, the probability 
distribution of finding $N$ electrons that have 
tunneled through the PC in time $t$.
The quantity $P(N,t)$ was discussed in terms of the ``partially'' 
reduced density matrix in Ref.\ \cite{Gurvitz98}
for a special case of the CQD/PC model, and it was analyzed in
Refs.\ \cite{Shnirman98}
and \cite{Makhlin01}
for a Cooper-pair-box charge qubit measured by a single-electron transistor. 
Here we present a detailed analysis of $P(N,t)$ for the CQD/PC model
using both of the ``partially'' reduced density
matrix and quantum trajectory approaches.
We show that the simulation results of $P(N,t)$
using $10000$ quantum trajectories and measurement records
are, as expected, in very good agreement with  
those obtained from the Fourier analysis of 
the ``partially'' reduced density matrix.
However, each single quantum trajectory and corresponding measurement record
mimics a single run of the measurement experiment.
Hence, even though we are interested in the time or ensemble average 
properties [such as $P(N,t)$] rather than conditional dynamics, 
the possible individual realizations of quantum trajectories 
and their corresponding measurement records do 
provide insight into, and aid in the
interpretation of, the average properties.
This appealing feature of the quantum trajectories is also illustrated in
this paper.

As just mentioned above, each single quantum trajectory and corresponding
measurement record resembles a possible single run of the 
continuous in time measurement experiment. 
We finally discuss, in this approach, the possibility of  
reading out the qubit state in a
single-shot measurement.
In particular, we clarify and provide conditions for 
single-shot readout of an initial charge qubit state.
Generally speaking, in order to obtain a confident CQD qubit state readout
in the charge state basis, it is 
better to (a) reduce the coherent coupling between the charge state,  
(b) increase the interaction with the PC detector
and (c) switch on the energy mismatch of the CQD's
for the readout measurement.

In Sec.~2, the CQD/PC model is briefly described and
the stochastic master equation or quantum trajectory equation is presented. 
In Sec.~3, it is shown that quantum trajectory theory
contains the most (all) information 
as far as the system evolution is concerned.
A close connection between the quantum
trajectory approach and master equation approach of the reduced
or ``partially'' reduced density matrix is established so that a unified
picture for these seemingly different approaches can be obtained.  
Then we demonstrate, further, 
this connection by considering
an experiment of reading out the initial CQD qubit state in 
Sec.~4. We show that the quantum trajectory approach not only give the
simulation results of $P(N,t)$ in an good agreement with those
obtained from the ``partially'' reduced density matrix approach, but
also provide more physical insight into the ensemble quantity $P(N,t)$.  
Then in Sec.~5, 
the issues of different time scales regarding to the 
quantum trajectory approach, 
the possibility of reading out the state of the
qubit in a single-shot experiment, and the possible experimental
implementation of the CQD/PC device are discussed.
Finally, a short conclusion is given in Sec.~6.

\section{Mesoscopic model and quantum trajectories}
\label{sec:QT}
\noindent
The CQD's measured by (interacting with) 
a low-transparency PC detector (bath)
has been extensively studied in Refs.\
\cite{Gurvitz97,Korotkov99,Makhlin00,Korotkov01b,Goan01a,Goan01b}.
Basically, when the electron
in the CQD system is near the PC (i.e., dot $1$ is occupied),
the effective energy independent 
tunneling amplitude 
of the PC detector
changes from $T_{00}\rightarrow T_{00}+\chi_{00}$ (see, e.g., Fig.~1
of Ref.\ \cite{Goan01a}).
As a consequence, the current through the PC is
also modified. This changed current can be detected,
and thus a measurement of 
the location of the electron in the CQD
system (qubit) is effected.
The unconditional and conditional master equation (or rate equations
for all the reduced-density-matrix elements) 
for the CQD system (qubit) 
has been
derived and analyzed 
in Refs.\ \cite{Gurvitz97,Goan01a} and 
Refs.\cite{Korotkov99,Korotkov01b,Goan01a,Goan01b}.
But the connection between the quantum trajectory (conditional,
stochastic master equation) approach and the master equation approach
of the ``partially'' reduced density matrix has not yet been formally
established. It is one of the major purposes of this paper to provide
such a connection between them, in terms of averaging over (tracing out) 
different amount of detection records (detector states).

For completeness,
the stochastic quantum-jump master equation of the density matrix operator,
conditioned on the observed event in the case of efficient
measurement in time $dt$ in Ref.\ \cite{Goan01a,Goan01b} is given
here again as:
\begin{eqnarray}
d\rho_c(t)
&=&dN_c(t)\left [\frac{{\cal J}[{\cal T}+{\cal X} n_1]}
{{\cal P}_{1c}(t)}
-1\right ]\rho_c(t)
\nonumber \\
&& +\, dt \left\{
-{\cal A}[{\cal T}+{\cal X} n_1]\rho_c(t)
+{\cal P}_{1c}(t) \rho_c(t)
-\frac{i}{\hbar}[{\cal H}_{CQD},\rho_c(t)] \right\}.
\label{condmasterEq}
\end{eqnarray}
The subscript $c$ indicates that the quantity to which 
it is attached is conditioned on previous measurement results.
$d\rho_c(t)=\rho_c(t+dt)-\rho_c(t)$ represents 
the change of the conditional density matrix operator in time $dt$.
${\cal H}_{\rm CQD}=\hbar [ \omega_1 c_1^\dagger c_1
+\omega_2c_2^\dagger c_2
+\Omega(c_1^\dagger c_2+ c_2^\dagger c_1)]$ 
represents the effective tunneling Hamiltonian
for the measured CQD system (charge qubit),
where $c_i$  ($c_i^\dagger$) and
$h\omega_i$ represent the electron annihilation (creation)
operator and energy for a single electron state in each dot
respectively, and the coherent coupling between these 
two dots is given by $h\Omega$.
The superoperators ${\cal J}$ and ${\cal A}$ are defined as
${\cal J}[B]\rho = B \rho B^\dagger$, and
${\cal A}[B]\rho = (B^\dagger B \rho +\rho B^\dagger B)/2$,
and $n_i=c_i^\dagger c_i$ is the
occupation number operator for dot $i=1,2$.
The parameters ${\cal T}$ and ${\cal X}$
are given by 
$D=|{\cal T}|^2= 2\pi e |T_{00}|^2 g_L g_R V_{\rm sd}/\hbar$, and
$D'=|{\cal T}+{\cal X}|^2=2\pi e |T_{00}+\chi_{00}|^2 g_L
g_R V_{\rm sd}/\hbar$.
Here $D$ and $D'$ are respectively the average electron tunneling rates
through the PC barrier without and with the presence
of the electron in dot $1$ (see, e.g., Fig.~1
of Ref.\ \cite{Goan01a}).
The external source-drain bias applied across the PC is given by
$eV_{\rm sd}=\mu_L-\mu_R$ 
($\mu_L$ and $\mu_R$ stand for the chemical potentials in the
left and right reservoirs respectively),
and $g_L$ and $g_R$ are the
energy-independent density of states for the left and right reservoirs.
In the {\it quantum-jump} case \cite{Goan01a},
the stochastic point processes $dN_c(t)$ in Eq.\
(\ref{condmasterEq}) represents
the number (either zero or one) of tunneling
events seen in an infinitesimal time $dt$ 
in the PC.
If no tunneling electron is detected, the result is {\em null}, 
i.e., $dN_c(t)=0$.
If there is {\em detection} of a tunneling electron in time $dt$, 
then $dN_c(t)=1$.
We can think of $dN_c(t)$ in Eq.\
(\ref{condmasterEq}) 
as the increment in the number of electrons
$N_c(t)=\sum dN_c(t)$ passing through PC barrier
in the infinitesimal time $dt$. 
It is the variable $N_c(t)$, the accumulated
electron number transmitted through the PC barrier,
which is used in Refs.\
\cite{Gurvitz97,Shnirman98,Makhlin01}.  
Since the nature of detection of 
electrons tunneling through the PC is stochastic, $dN_c(t)$ 
should represent a classical random process. 
In addition, the ensemble
average $E[dN_c(t)]$ of the classical stochastic process $dN_c(t)$ 
should equal
the probability (quantum average) of
detecting electrons tunneling through the PC barrier in time $dt$.
Hence, we have (see, e.g., Eqs.\ (22), (29) and (30) of Ref.\ \cite{Goan01a}) 
\begin{equation}
E[dN_c(t)] = {\cal P}_{1c}(t) dt,
\label{dNc}
\end{equation}
where
\begin{equation}
{\cal P}_{1c}(t)= D+(D'-D)\langle n_1\rangle_c(t),
\label{Plc}
\end{equation}
$\langle n_1\rangle_c(t)={\rm Tr}[n_1\rho_c(t)]$,
and $E[Y]$ denotes an ensemble
average of a classical stochastic process $Y$.
Formally we can write the current through the PC as 
$I_c(t) = e\, {dN_c(t)}/{dt}$. 
It is then easy to see that Eqs.\ (\ref{dNc}) and (\ref{Plc})
simply state that when dot $1$ is empty 
the average current through the PC is $eD$. 
While it is $eD'$ when dot 1 is occupied.
In the quantum trajectory approach, 
the instantaneous system state $\rho_c(t)$
conditions the measured
current [see Eqs.\ (\ref{dNc}) and (\ref{Plc})], 
while the measured current [$dN_c(t)$ in Eq.\ (\ref{condmasterEq})]  
conditions the future evolution of the measured system
in a self-consistent manner.

\section{Connections to master equation approach}
\label{sec:masterEq}
\noindent
We show next that quantum trajectory theory or conditional, stochastic 
density matrix 
contains the most (all) information 
as far as the system evolution is concerned,
and the master equation for the 
reduced or ``partially'' reduced density matrix simply results when an
average or ``partial'' average is taken on the conditional, stochastic 
master equation (\ref{condmasterEq}) over the possible outcomes of the
measurements on the PC detector.
This provides a unified picture, 
in terms of averaging over (tracing out) 
different amount of detection records (detector states),
for these seemingly different approaches reported in the literature.

The traditional, unconditional master equation approach is to
define a total system-environment density matrix. 
Then tracing over the
states of the environments (PC detectors) 
leads to the reduced density matrix
for the CQD qubit system alone.  
The effect of integrating or tracing out the
environmental (detector)
degrees of the freedom to obtain the reduced density matrix is
equivalent to that of completely ignoring or averaging over the
results of all measurement records $dN_c(t)$. 
Hence the unconditional master
equation can be obtained as in  Refs.\ \cite{Goan01a} and 
\cite{Goan01b} by 
taking the ensemble average over
the observed stochastic process on  
Eq.\ (\ref{condmasterEq}).
By setting $E[dN_c(t)]$ equal to its 
expected value Eq.\ (\ref{dNc}), and replacing
$E[\rho_c(t)]=\rho(t)$ in Eq.\ (\ref{condmasterEq}), 
we then find the resultant unconditional master equation as
\begin{equation}
\dot{\rho}(t)=-\frac{i}{\hbar}[{\cal H}_{\rm CQD}, \rho(t)]
+{\cal J}[{\cal T}+{\cal X} n_1]\rho(t)
-{\cal A}[{\cal T}+{\cal X} n_1]\rho(t)
\label{masterEq}
\end{equation}
where $\dot{\rho}(t)=d\rho(t)/dt$.
In this approach of master equation of the reduced density matrix, 
the influence of the PC environments on the CQD system can be
analyzed \cite{Goan01a,Gurvitz97}. 
For example, the total decoherence or
dephasing rate due to the PC environment (detector) 
can be found to be $\Gamma_d=|{\cal X}|^2/2$.
But this approach or Eq.\ (\ref{masterEq})
does not tell us anything about the 
experimental observed quantity, namely the electron counts or current 
through the PC. Hence, the PC detector in this approach are treated as
a pure environment for the system, rather than a measurement device.

An alternative approach recently developed in 
Refs.\ \cite{Gurvitz97,Shnirman98,Gurvitz98,Makhlin00,Makhlin01}
is to take trace over
environmental (detector) microscopic degrees of the freedom but keep track the
number of electrons, $N$, that have tunneled through the
PC barriers during time $t$ in the reduced density matrix. 
This allows one to extract information about the quantum state of the
qubit, by measuring electron number counts $N$ 
through the PC in time $t$. This approach is regarded here as the
``partially'' reduced density matrix approach. 
The master (rate) equation of the ``partially'' 
reduced density matrix for the
CQD qubit system measured by a PC detector 
was derived in Refs.\ \cite{Gurvitz97} and \cite{Gurvitz98} 
from the so-called many-body Schr\"odinger
equation. While it was derived in 
Refs.\ \cite{Shnirman98,Makhlin00} and \cite{Makhlin01}, 
by means of the diagrammatic technique
in the Keldysh forward and backward in time contour,
for a Cooper-pair-box charge qubit coupled
capacitively to a single-electron transistor detector.      
Here we show that it can be obtained 
for the CQD/PC model by simply taking a
``partial'' average on the conditional, stochastic 
master equation (\ref{condmasterEq})
of the CQD qubit system 
density matrix over the possible outcomes of the
measurements of the PC detectors.
For the case of {\em quantum jumps}, \cite{Goan01a} there are
two possible measurement outcomes of the PC detector in time $dt$, 
namely {\em null} (no electron detected) and {\em detection} of an
electron passing through the PC barrier.
As discussed in Ref.\ \cite{Goan01a,Wiseman96} and
\cite{Goan02},  the effect of
the {\em detection} of an electron passing through the PC barrier in
time interval [$t$,$t+dt$) on the CQD qubit density matrix
is described by the ${\cal J}$ superoperator term 
(see, e.g., Eq.\ (29) of Ref.\ \cite{Goan01a}).   
This is why sometimes ${\cal J}$ is called a {\em jump} superoperator.
The procedure to take the ``partial'' average can now be described as follows.
First, taking the ensemble average on Eq.\ (\ref{condmasterEq}), we
obtain Eq.\ (\ref{masterEq}). Then in order to keep track of the
number of electrons $N$ that have tunneled, 
we need to identify the effect of the
{\em jump} superoperator ${\cal J}$ term in Eq.\ (\ref{masterEq}).
If $N$ electrons have
tunneled through the PC at time $t+dt$, then the accumulated number
of electrons in the drain of the PC at the earlier time $t$, 
due to the contribution of the {\em jump} term of the PC,
should be $(N-1)$.
Hence, after writing out the number dependence $N$ or $(N-1)$
explicitly for the density matrix in Eq.\ (\ref{masterEq}), 
we obtain the master equation for the ``partially'' reduced density
matrix as:
\begin{eqnarray}
\dot{\rho}(N,t)&=&-\frac{i}{\hbar}[{\cal H}_{\rm CQD}, \rho(N,t)]
+{\cal J}[{\cal T}_u+{\cal X}_u n_1]\rho(N-1,t)
-{\cal A}[{\cal T}_u+{\cal X}_u n_1]\rho(N,t)
\label{masterEqN}
\end{eqnarray}
Note that the index $(N-1)$ appearing in the {\em jump} superoperator
${\cal J}$ terms of Eq.\ (\ref{masterEqN}).
If the sum over all possible values of $N$ is
taken on the ``partially'' reduced density matrix [i.e., tracing out
the detector states completely,  
$\rho(t)=\sum_{N} \rho(N,t)$], 
Eq.\ (\ref{masterEqN}) then reduces to Eq.\ (\ref{masterEq}).
This procedure of reducing Eq.\ (\ref{condmasterEq}) to 
Eq.\ (\ref{masterEqN})
and then to Eq.\ (\ref{masterEq}) by tracing out more and more 
available detector information
is one of the main results of the paper.
This procedure is particularly simple if  
the master equations are
expressed in a form in terms of superoperators ${\cal J}$ 
and ${\cal A}$, and the effect of the {\em jump} superoperator 
${\cal J}$ term is identified.

In order to see our approach does reproduce the result in 
Refs.\ \cite{Gurvitz97} and \cite{Gurvitz98},
we consider the case of real tunneling amplitudes for the PC.
In this case the relative phase between the two tunneling amplitudes 
$\theta=\pi$. This is because the presence of the electron of the CQD
system near the PC detector raises the effective tunneling barrier
of the detector, so the average electron tunneling rate through the PC,   
$D'=|{\cal T}+{\cal X}|^2 < D=|{\cal T}|^2$.
After evaluating Eq.\ (\ref{masterEqN}) in the logical qubit charge state
$|a\rangle$ and $|b\rangle$ (i.e., perfect localization state of the charge 
in dot 1 and dot 2, respectively), 
we obtain the rate equations as:
\begin{eqnarray}
\dot\rho_{aa}(N,t) & = & i\Omega [\rho_{ab}(N,t)-\rho_{ba}(N,t)]
+|{\cal T}+{\cal X}|^2\rho_{aa}(N-1,t)-|{\cal T}+{\cal X}|^2\rho_{aa}(N,t)\;,
\label{rateEqa}\\
\dot\rho_{bb}(N,t) & = & i\Omega [\rho_{ba}(N,t)-\rho_{ab}(N,t)]
+|{\cal T}|^2\rho_{bb}(N-1,t)-|{\cal T}|^2\rho_{bb}(N,t)\;,
\label{rateEqb}\\
\dot\rho_{ab}(N,t) & = & i{\cal E}\rho_{ab}(N,t)+
i\Omega [\rho_{aa}(N,t)-\rho_{bb}(N,t)]
+(|{\cal T}|\; |{\cal T}+{\cal X}|) \rho_{ab}(N-1,t)
\nonumber \\
&&-\frac{1}{2}(|{\cal T}+{\cal X}|^2+|{\cal
T}|^2)\rho_{ab}(N,t)\, ,
\label{rateEqc}
\end{eqnarray}
where $\rho_{ij}=\langle i|\rho |j\rangle$ with $i,j=a,b$ refer to the
qubit charge basis, and 
$\hbar {\cal E}=\hbar (\omega_2-\omega_1)$ is the energy
mismatch between the two dots. 
By setting $|{\cal T}|^2=D$, $|{\cal T}+{\cal X}|^2=D'$ and 
$(|{\cal T}| \; |{\cal T}+{\cal X}|)=\sqrt{D \; D'}$, we find that
Eqs.\ (\ref{rateEqa})--(\ref{rateEqc}) are the same as 
Eq.(3.3) of Ref.\ \cite{Gurvitz97}.
While the ``partially'' 
reduced density matrix approach provides us with information about 
the number of accumulated electrons
passing through the PC, 
it is still in an ensemble and time average sense. 
In other words, the system dynamics is still deterministic in this approach.
Hence, it cannot describe
the conditional dynamics of the CQD qubit system in a single realization of
continuous measurements, which reflects the stochastic nature of
electrons tunneling through the PC barrier.

On the other hand, in the quantum trajectory or stochastic
Schr\"odinger equation approach, 
no average or trace over the bath states is
taken as far as the system evolution is concerned. 
Instead, repeated continuous in time measurements are made, and
the measurement results are recorded. So in this approach, we are  
propagating {\em in parallel} the information of a conditioned
(stochastic) state evolution $|\psi_c(t)\rangle$ 
[or equivalently a conditioned density matrix evolution $\rho_c(t)$]
and a detection record 
$dN_c(t)$ in a single run of a
continuous measurement process.
The stochastic element in the quantum trajectory corresponds exactly
to the consequence of the random outcomes of the detection record.

In summary, 
the master equations of the
reduced or ``partially'' reduced density matrix can be obtained  
as a result of taking an ensemble average or partial average over the possible
measurement records in the quantum trajectory approach.
This procedure of reducing Eq.\ (\ref{condmasterEq}) to 
Eq.\ (\ref{masterEqN})
and then to Eq.\ (\ref{masterEq}),
by averaging over (tracing out) more and more information of the 
detection records (detector states),
provides us with a unified picture 
for these seemingly different approaches.
If only one measurement value is recorded in
each run of experiments (for example, the number of electrons $N$ that have
tunneled in time $t$) and
ensemble average properties 
[for example, $P(N,t)$, see Sec.\ \ref{sec:readout}] are studied
over many repeated experiments, the quantum trajectory
approach will give the same result as
the master equation approach of the reduced or ``partially'' reduced 
density matrix.
However, more physical insights in the
interpretations of the ensemble average properties
can be gained in terms of 
different realizations of quantum trajectories and 
their corresponding measurement records.
We will discuss this appealing feature of the quantum trajectory approach
in next section by considering an example of a readout measurement
experiment of the initial CQD qubit state.

\section{Reading out the initial qubit state}
\label{sec:readout}
\noindent
To read out the initial qubit state $\rho_{aa}(0)=|a|^2$ and 
$\rho_{bb}(0)=|b|^2=1-|a|^2$,
we may define, as in Refs.\
\cite{Shnirman98,Makhlin00} and \cite{Makhlin01}, 
the probability distribution of finding $N$
electrons that have tunneled during time $t$: 
\begin{equation}
P(N,t)={\rm Tr}_S[\rho(N,t)],
\label{PNt}
\end{equation} 
where ${\rm Tr}_S$ means tracing the density matrix over the degrees
of freedom of the qubit system.
The quantity $P(N,t)$ has been analyzed in terms of the ``partially'' 
reduced density matrix in Refs.\ \cite{Shnirman98,Makhlin00}
and \cite{Makhlin01}
for a Cooper-pair-box charge qubit coupled
capacitively to a single-electron transistor. 
Here, to demonstrate the connection between quantum trajectory and
``partially'' reduced density matrix approaches,
we present the case for the CQD/PC model first using the 
same method and then the quantum trajectory approach.

\subsection{Partially reduced density matrix approach}
\label{subsec:PRDM}
\begin{figure}[htbp]
\centerline{\psfig{file=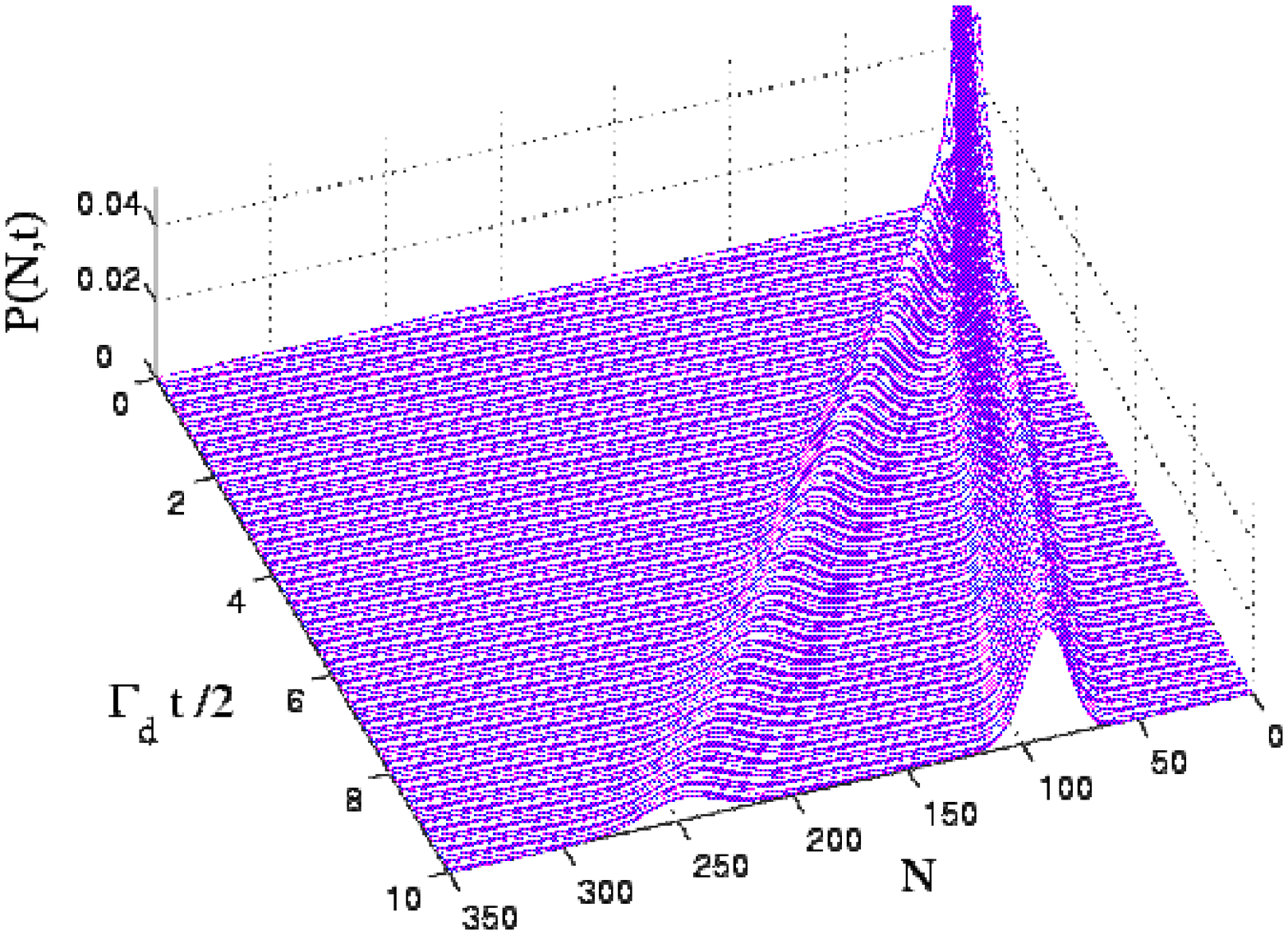,width=0.76\linewidth,angle=0}}
\fcaption{Probability distribution of finding $N$
electrons that have tunneled through the PC
during time $t$, $P(N,t)$, plotted vs $t$ and $N$ with 
the initial qubit state being $\sqrt{3/4}|a\rangle+(1/2)|b\rangle$.
Other parameters are: ${\cal E}=0$, $\theta=\pi$, 
$|{\cal T}|^2=25 |{\cal X}|^2/4=25 \Gamma_d/2=250 \Omega$,
and time is in unit of $2 (\Gamma_d)^{-1}$.
In the plot, in order to have a clear contrast for the distribution
in the specified time region, 
the range of $P(N,t)$ values is limited to be smaller than $0.05$.  
Note that $P(N,0)=\delta_{N,0}$.}
\label{fig:PNt3d}
\end{figure}

\noindent
The quantity $P(N,t)$, defined in Eq.\ (\ref{PNt}) in
the CQD charge basis, can be written as 
$P(N,t)=\rho_{aa}(N,t)+\rho_{bb}(N,t)$.
To obtain the solution of $\rho_{ij}(N,t)$ with $i,j=a,b$
in the ``partially'' reduced density matrix approach, 
we can first apply a
Fourier transform \cite{Shnirman98,Gurvitz98,Makhlin00,Makhlin01} 
$\rho_{ij}(k,t)=\sum_N e^{-ikN}\rho_{ij}(N,t)$ to 
Eqs.\ (\ref{rateEqa})--(\ref{rateEqc})
since these equations are translationally invariant in $N$ space.
Solving the resultant coupled first-order differential equations for
$\rho_{ij}(k,t)$ and then performing a inverse Fourier 
transform, we obtain the probability distribution 
$P(N,t)=\rho_{aa}(N,t)+\rho_{bb}(N,t)$.  
The result is illustrated in Fig.\ \ref{fig:PNt3d}.
At $t=0$, no electrons have tunneled so $P(N,0)=\delta_{N,0}$. 
Then the probability distribution (for parameters in the Zeno regime, 
discussed later)
moves to positive $N$ values with two different velocities (slopes in the $N$,
$t$ space) $D'$ and $D$, which correspond to the average electron
tunneling rates through PC 
for CQD qubit state being in $|a\rangle$ and $b\rangle$ states, respectively.
Simultaneously, the width, $\sqrt{2D't}$ or $\sqrt{2Dt}$, of the respective
distribution due to shot noise widens with time. 
Due to this intrinsic shot noise on the measurement output, 
the readout of the initial
qubit state necessarily needs finite time duration in order to have 
an acceptable signal-to-noise ratio. 
So if after some time 
(larger than the measurement time $t_{\rm m}$ or 
localization time \cite{Goan01a} $t_{\rm loc}$), 
the two distributions become
sufficiently separated and distinguishable, and their weights still
correspond closely to the
initial values of the diagonal elements of the qubit density matrix in
the qubit charge basis, then the initial qubit state probability can be
read out. 
After a longer time (the mixing time $t_{\rm mix}$) due to
transitions between the two charge states, taken into account the
back action of the measurements on the qubit, 
the valley of the two distributions would
fill up and a single broad plateau would develop.
Hence there is an optimal time window $t_{\rm m}\ll t\ll t_{\rm mix}$ 
for which a confident readout measurement can be performed.  

\begin{figure}[htbp]
\centerline{\psfig{file=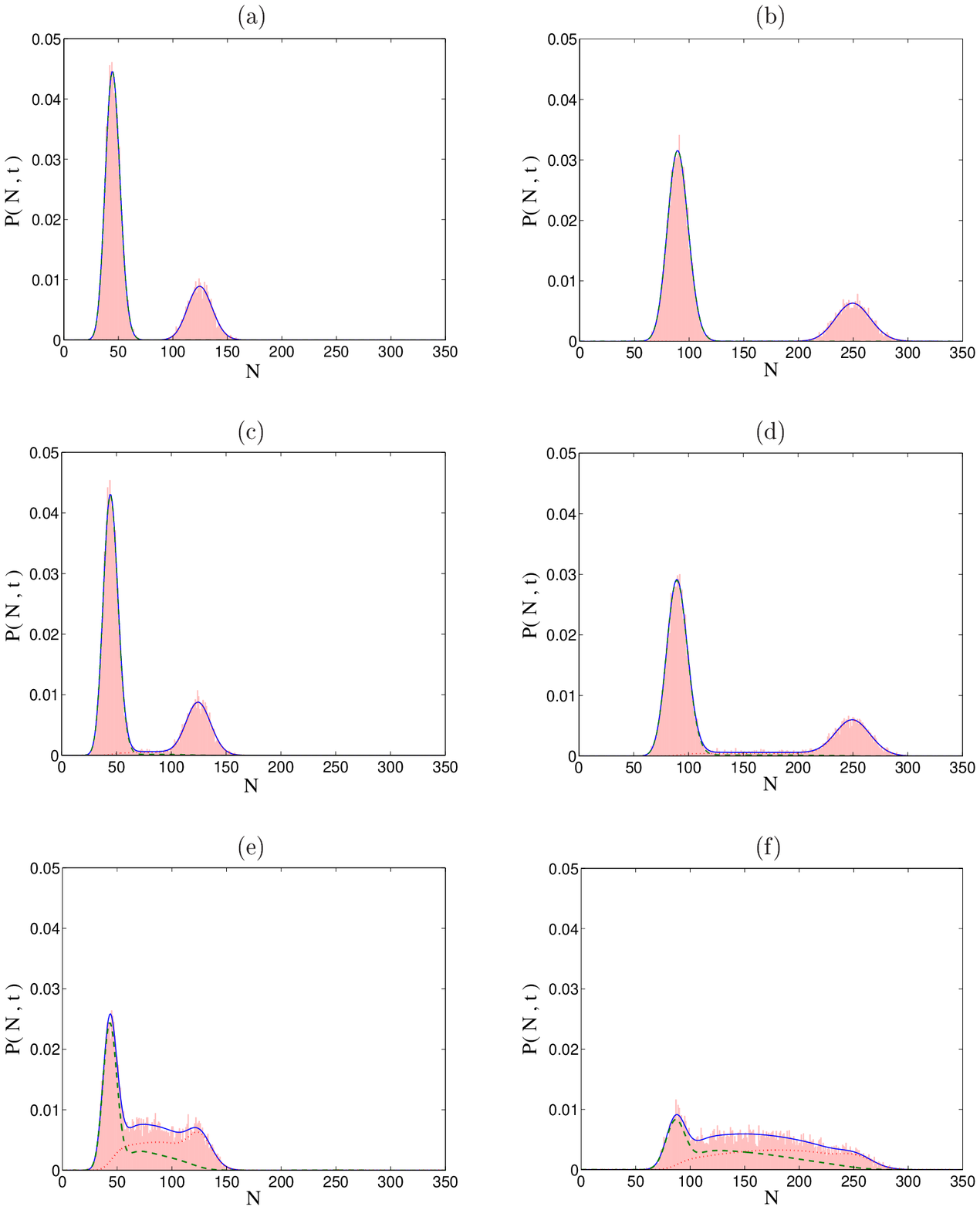,width=0.93\linewidth,angle=0}}
\fcaption{Probability distribution $P(N,t)$ for a fixed value of $\Gamma_d$
but with different ratios of $(\Omega/\Gamma_d)=0,0.05,0.2$. The
initial amplitudes of the qubit state are $\rho_{aa}(0)=|a|^2=0.75$
and $\rho_{bb}(0)=|b|^2=0.25$. 
The ratios of $(\Omega/\Gamma_d)$ 
for (a) and (b) is $0$, for (c) and (d) is $0.05$, and for (e) and (f)
is $0.2$.
The detection time (in unit of $2 (\Gamma_d)^{-1}$) 
for (a), (c) and (e) is up to $t=10/\Gamma_d$ , and 
for (b), (d) and (f) is to $t=20/\Gamma_d$.
Other parameters are: 
${\cal E}=0$, $\theta=\pi$, $|{\cal T}|^2=25 |{\cal X}|^2/4=25 \Gamma_d/2$.
The quantities $\rho_{aa}(N,t)$, $\rho_{bb}(N,t)$ and 
$P(N,t)=\rho_{aa}(N,t)+\rho_{bb}(N,t)$ obtained from the 
Fourier analysis are represented in each plot as dashed, 
dotted, and solid line, respectively.
The shaded region in each plot is the simulation result of using 
$10000$ quantum trajectories and their corresponding detection records.
The expected average number of electron counts $Dt$ and $D't$ 
of the two peaks 
for (a), (c) and (e) with detection time up to $t=10/\Gamma_d$
is $45$ and $125$ respectively, and it is $90$ and $250$ 
for (b), (d) and (f) with $t=20/\Gamma_d$.}
\label{fig:PNt}
\end{figure}

Of course, the
measurement of the PC, which results in a back action on the qubit system, 
will influence the qubit system evolution.
For fixed ${\cal E}$, ${\cal T}$, and $\Omega$,
increasing the value of ${\cal X}=\sqrt{2 \Gamma_d}$, we increase the
interaction with the PC. 
As shown for typical conditional evolutions or quantum
trajectories for symmetric 
(i.e., ${\cal E}=0$)  CQD's in Fig.\ 4 of Ref.\ \cite{Goan01b},  the
period of coherent oscillations between the two CQD qubit states
increases with increasing $(\Gamma_d/\Omega)$,
while the time of a transition (switching time) decreases.
Now in order to read out the initial CQD qubit state, we want 
the qubit during the detection process to stay in one of 
its charge states much longer than $t_{\rm m} \sim t_{\rm loc}$, 
but much smaller than the average time between
state-changing transitions (the mixing time $t_{\rm mix}$), 
i.e., $t_{\rm m} \ll t\ll t_{\rm mix}$. 
It was shown \cite{Goan01a,Shnirman98} that 
in the regime of small $(\Omega/\Gamma_d)$ ratio \cite{Goan01a,Shnirman98}, 
the measurement (localization) 
time $t_{\rm m} \sim t_{\rm loc} \approx (1/\Gamma_d)$ 
is much smaller than the mixing time
$t_{\rm mix} \approx (\Gamma_d^2+{\cal E}^2)/(4\, \Omega^2 \, \Gamma_d)$.
This is the (Zeno) regime that we will consider.
Figure \ref{fig:PNt} 
shows the probability distributions $P(N,t)$ of 
symmetric CQD's for
a fixed value of $\Gamma_d$
but with different ratios of $(\Omega/\Gamma_d)$. 
The detection time intervals are up to $t=10/\Gamma_d$, or 
$t=20/\Gamma_d$.
We can see from Fig.\ \ref{fig:PNt}(a)-(d)
that for smaller ratio of $(\Omega/\Gamma_d)=0,0.05$,
the distribution splits into two and their 
weights corresponds closely to the initial values of the diagonal elements
of $\rho_{aa}(0)=0.75$ and $\rho_{bb}(0)=0.25$
[the latter is supported by the fact that the ensemble average $\rho_{aa}(t)$
deviates from $\rho_{aa}(0)=0.75$ only a little bit for all times 
within $t=20/\Gamma_d$].
In this case, a good quantum readout 
measurement for the CQD qubit initial state
can be performed by repeatedly measuring $N(t)$ in these time durations.
For larger ratio of $\Omega/\Gamma_d=0.2$, 
the valley between the two-peak structure
is already partially filled for $t=10/\Gamma_d$, as shown in 
Fig.\ \ref{fig:PNt}(e). 
After a longer time at $t=20/\Gamma_d$, 
the two peaks evolve into a broad plateau (see Fig.\ \ref{fig:PNt}(f)). 
This indicates that (on average) 
mixing transitions take place on the same time
scale as that of peak separation.
Therefore, no good quantum measurement of the qubit initial state 
can be performed in this case.

\begin{figure}[htbp]
\centerline{\psfig{file=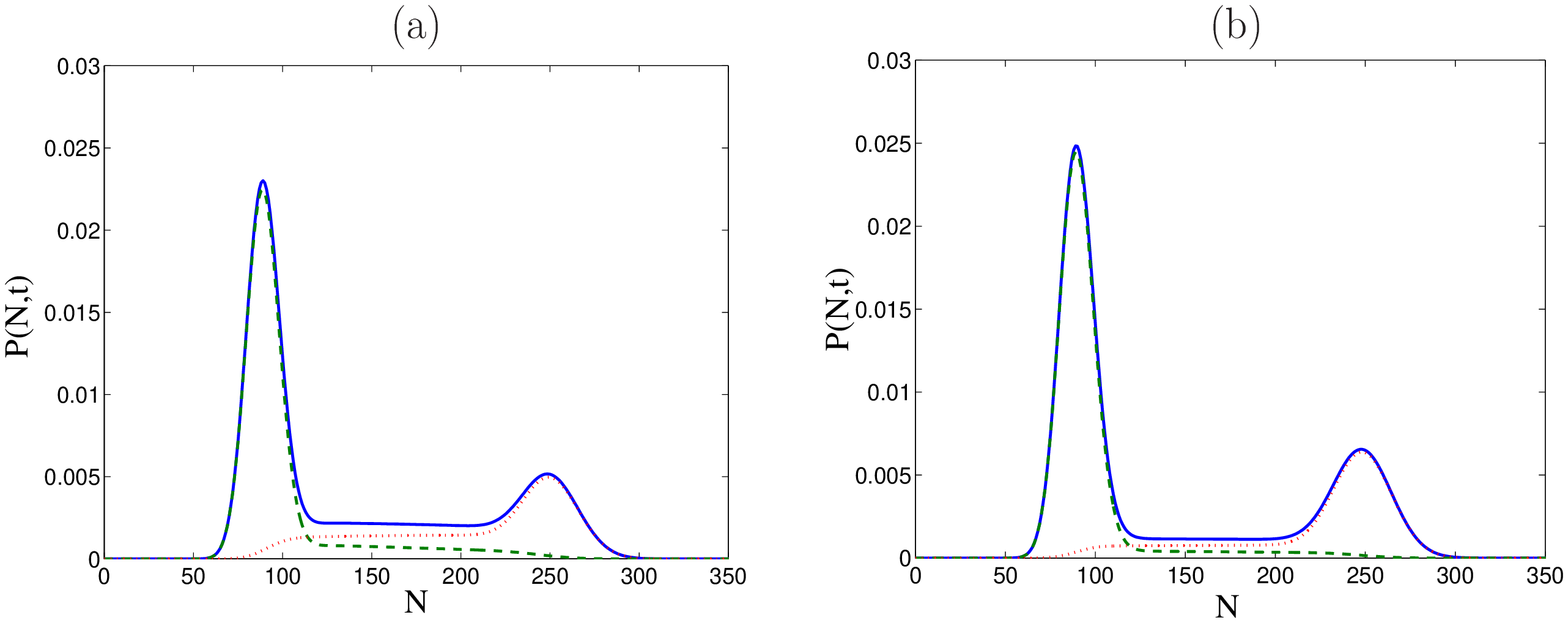,width=5.8cm,angle=90}}
\fcaption{The quantities $\rho_{aa}(N,t)$, $\rho_{bb}(N,t)$ and 
$P(N,t)=\rho_{aa}(N,t)+\rho_{bb}(N,t)$ obtained from the 
Fourier analysis of the ``partially'' reduced density matrix,
and represented as dashed, 
dotted, and solid line, respectively, in each plot
for detection time interval $t=20/\Gamma_d$: 
(a) symmetric CQD's with
${\cal E}=0$ and (b) asymmetric CQD's with ${\cal E}=\Gamma_d$.
Other parameters are: $\theta=\pi$,
$|{\cal T}|^2=25 |{\cal X}|^2/4=25 \Gamma_d/2=125 \Omega$.}
\label{fig:compare}
\end{figure}

The effect of energy mismatch ${\cal E}$ of asymmetric CQD's on the
$P(N,t)$ is that the mixing rate $\gamma_{\rm mix}=t^{-1}_{\rm mix}$ 
is smaller than that of symmetric CQD's. This can be observed 
in Fig.\ \ref{fig:compare}. 
The weight in the overlap region of $P(N,t)$ 
in Fig.\ \ref{fig:compare}(b) for asymmetric CQD's
is smaller than that in Fig.\ \ref{fig:compare}(a) for symmetric CQD's
up to the same detection time interval $t=20/\Gamma_d$. 
This suggests that if the CQD qubit undergoes additional coherent
evolution [i.e., $\Omega \neq 0$, the source of mixing behavior in $P(N,t)$]
during the readout process of its initial state, it
is better to switch on the energy mismatch ${\cal E} \neq 0$ of the CQD's
right before the readout measurement as in the case of 
Cooper-pair-box charge qubit in Ref.\ \cite{Nakamura99}.

\subsection{Quantum trajectory approach}
\label{subsec:QT}
\noindent
In the approach of quantum trajectories, more physical insights can, 
however, be gained. 
Each single trajectory resembles a single history of the system state 
in a single run of a continuous in time measurement experiment.
We can therefore 
use quantum trajectories and their corresponding detection records 
to simulate measurement experiments on a single quantum system.  
For example, the quantity $P(N,t)$ defined in Eq.\ (\ref{PNt}) can be
simulated by building the histogram of the accumulated number of
electrons $N_c(t)=\sum dN_c(t)$ up to time $t$ for many realizations of
the detection records (generated together with their corresponding 
quantum trajectories), and then normalizing the distribution to one.
Figure \ref{fig:PNt} shows that the probability distributions 
$P(N,t)$, each constructed from $10000$ possible realizations (in shades), 
are, as expected, in very good
agreement with those obtained from the Fourier analysis of 
the partially reduced density matrix (in solid lines).

\begin{figure}[htbp]
\centerline{\psfig{file=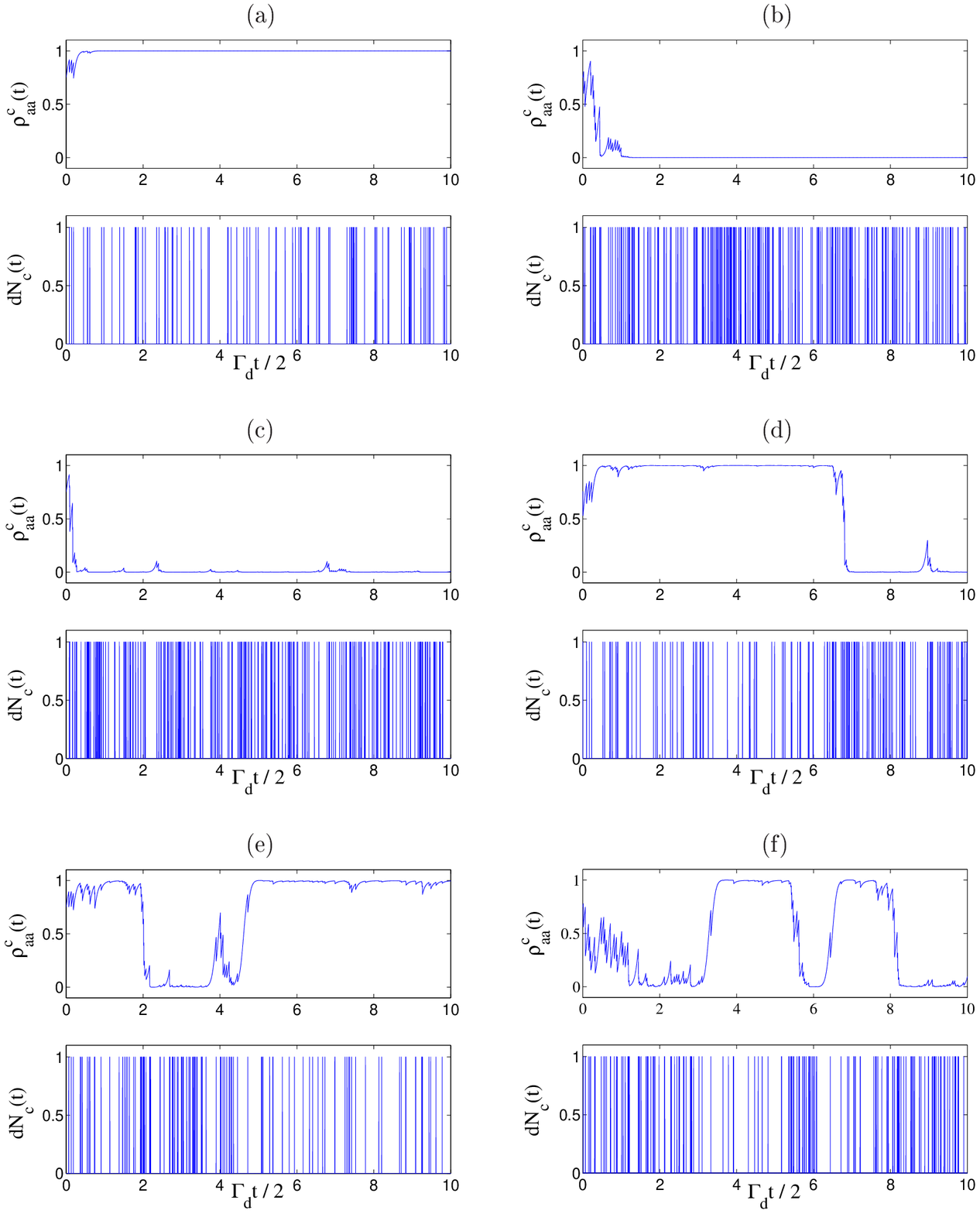,width=\linewidth,angle=0}}
\fcaption{Typical quantum trajectories and corresponding detection
records for $P(N,t)$ shown in Fig.\ \ref{fig:PNt}.
The ratios of $(\Omega/\Gamma_d)$ 
for (a) and (b) is $0$, for (c) and (d) is $0.05$, and for (e) and (f)
is $0.2$.
Other parameters are the same as in Fig.\ \ref{fig:PNt}.
The number of electron $N_c(t)=\sum dN_c(t)$ 
that have tunneled through PC up to time 
$t=10/\Gamma_d$ and $t=20/\Gamma_d$ is respectively $43$ and $93$ in
(a), $132$ and $237$ in (b), $129$ and $248$ in (c),  
$50$ and $157$ in (d), $83$ and $122$ in (e),
and $75$ and $168$ in (f).}   
\label{fig:trajectories}
\end{figure}

Note that the time scales $t_{\rm m}$ (or $t_{\rm loc}$) and
$t_{\rm mix}$ are basically statistical ensemble average quantities.
We can gain more physical insight, for example, on the occurrence of the
mixing, i.e., the occurrence of the overlap region between the
two distributions in $P(N,t)$.   
The fundamental origin of the mixing is due to the presence of the 
non-zero coherent coupling term between two preferred basis states 
of the qubit in the Hamiltonian. 
In our case, it is the non-zero $\Omega$ term which causes transitions
between the two qubit charge states. 
For $\Omega=0$, the conditional state
becomes localized in one of the two
dots, 
i.e., either $\rho^c_{aa}(t)=1$ or $\rho^c_{aa}(t)=0$ 
[see typical quantum trajectories in
Fig.\ \ref{fig:trajectories}(a) and (b)]. Hence
no overlap between the two distributions in $P(N,t)$ occurs for
$t>t_{\rm m}$ [see Fig.\ \ref{fig:PNt}(a) and (b)].
Besides, it can be easily shown that when $\Omega=0$ the ensemble averaged
$\rho_{aa}(t)$ and $\rho_{bb}(t)$ are constants of motion, i.e., 
$\rho_{aa}(t)=\rho_{aa}(0)$ and $\rho_{bb}(t)=\rho_{bb}(0)$. 
However, when $\Omega\neq 0$, the overlap
between the two distributions develops, and the weight in the overlap
region increases with increasing $(\Omega/\Gamma_d)$ ratio
as illustrated in Fig.\ \ref{fig:PNt}. 
The typical quantum trajectories whose accumulated electron detection
number falls in the peak regions are shown in 
Fig.\ \ref{fig:trajectories}(a)-(c). 
Those in overlap regions are shown in 
Fig.\ \ref{fig:trajectories}(d)-(f).
Note that the accumulated detected electron number of 
the first half of the time evolution
in Fig.\ \ref{fig:trajectories}(d) 
(i.e., for detection time up to $t=10/\Gamma_d$), actually 
falls in the peak region of $\rho_{aa}(N,t)$. 
Within the readout time $t$, 
if the qubit changes its state, then its
accumulated electron number is very probably falling into the valley 
region between the peaks of the two distributions. 
For fixed $\Gamma_d$, as the values of $\Omega$ increase,
the chance for the state-changing transitions to happen in each
individual realization of the quantum trajectory increases. If
the ratio $(\Omega/\Gamma_d)$ is increased further, 
the number of state-changing
transitions within the same readout time interval also increase
[see, for example, Fig.\ \ref{fig:trajectories}(e) and (f)].
As a result, more weight of the distributions move to fill the
valley and the two peaks transform into a broad plateau
[see Fig.\ \ref{fig:PNt}(e) and (f)]. 
These typical quantum trajectories 
indeed help us understand the occurrence of the mixing.

In summary, many different individual realizations of
quantum trajectories and measurement records 
provide more information and 
do give physical insight into, 
and aid in the interpretation of, the ensemble average properties,
for example, the quantity $P(N,t)$ discussed here.

\section{Discussions}
\label{sec:discussion}

\subsection{Different time scales}
\label{subsec:timescales}
\noindent
Next we discuss briefly the relationship between different 
time scales involved:
(a) the required detection time interval $dt$ for the quantum trajectory
equations to be valid, 
(b) the actual detection time interval, ($1/\Delta \omega$), where
$\Delta \omega$ is the bandwidth of the measurement device.
(c) the PC bath correlation time $\tau_B\approx \hbar/eV_{\rm sd}$,
(d) the typical system evolution or response time 
$t_{\rm res}=\min(1/\Omega, 1/{\cal E},1/\Gamma_d)$, 
where $\tau_d=1/\Gamma_d$ is known as the decoherence (dephasing) time
in the reduced density matrix approach, 
(e) the typical time $\min(1/D,1/D')$ for an individual 
electron tunneling through the PC,
(f) the measurement time $t_{\rm m}$ for being able to distinguish the two
probability distributions in $P(N,t)$,
(g) the mixing time $t_{\rm mix}$ after which the two distributions in
$P(N,t)$ transform into a single, broad distribution so that the
information about the initial qubit state is lost.    

The detection time interval $dt$ for the quantum trajectory approach 
should be much smaller than
$t_{\rm res}$ so that effectively the system is under continuous measurement. 
But it cannot be arbitrarily small.
This is because we should allow the PC baths
sufficient time to relax. 
If not, then the
buildup of PC bath electron population in other number (Fock) state 
will be suppressed by the continuous measurement,
and one would encounter the
situation analogous to the quantum Zeno effect. 
For the Born-Markov quantum-jump 
stochastic master equation\cite{Goan01a,Goan01b} to be valid,
we should require 
$\tau_B \ll dt=(1/\Delta \omega) \leq \min(1/D,1/D') \ll t_{\rm res}$, 
and for the quantum-diffusive case \cite{Korotkov99,Goan01a,Goan01b}, 
$\tau_B \ll  \min(1/D,1/D') \ll dt=(1/\Delta \omega)  \ll t_{\rm res}$. 

\subsection{Single-shot readout}
\label{subsec:singleshot}
\noindent
One of the interesting questions in the measurement experiment is whether
it is possible to read out the qubit state in a single-shot measurement. 
We shall discuss this question using quantum trajectory approach 
for the following cases: (I) to read
out the qubit state, $\rho_{aa}(t)$, at the detection time $t$, (II) to read
out the initial qubit state, $\rho_{aa}(0)$, (III) to read out the 
two possible charge states of a classical charge signal 
undergoing random jumps,  
or to observe the incoherent charge transfer between two quantum dots. 
It was stated \cite{Aassime01,Johansson02}
that if the ratio between the mixing time and the
measurement time is much larger than unity $(t_{\rm mix}/t_{\rm m})\gg 1$, 
it should be possible to
read out the qubit state in a single measurement. 
Since the time scales $t_{\rm m}$ and
$t_{\rm mix}$ are basically statistical ensemble average quantities,
strictly speaking the single-shot readout is possible for case I
only when $(t_{\rm mix}/t_{\rm m})\to \infty$. 
For case II, it is possible provided that  
$(t_{\rm mix}/t_{\rm m})\to \infty$, and 
$\rho_{aa}(0)=1$ or $\rho_{aa}(0)=0$, i.e.,  
the initial qubit state is in one of the
two perfectly localized states.
In practice, provided certain required measurement fidelity threshold
is met, one may relax the condition 
$(t_{\rm mix}/t_{\rm m})\to \infty$ for cases I and II
to $(t_{\rm mix}/t_{\rm m})\gg 1$. 
For case III, it is possible as long as the largest frequency for the charge
state change is smaller than the bandwidth of the detector.
We discuss further these three cases in the following.

Case I: 
If the bandwidth and other time scales satisfy the requirement for the
quantum trajectory equations to be valid, then given that the initial
state is known, we can, in principle, know the state of the qubit conditioned
on the measurement record after each detection time interval
in each single run of experiment.
If the bandwidth is not large enough $\Delta \omega < (1/t_{\rm res})$,
the measurement device would perform
a time average over the time interval $(1/\Delta \omega)$,
which is larger than the typical system evolution or response
time $t_{\rm res}$. In this case, 
even with the detection record from the measurement
device, we lose some information about the changes of the qubit
state. The decrease in our knowledge of the qubit state would then lead to 
decoherence (dephasing) for the qubit state. 
In other words, the conditional stochastic master equation 
would not be valid, 
and we do not know for sure the state of the qubit after each
detection time interval.

In the present CQD/PC model,
when $\Omega=0$, the CQD qubit will eventually 
(typically after $t>t_{\rm m}\sim t_{\rm loc}$)
localize in one of the two dots and stay there
(fixed points).
One can then determine the qubit state, $\rho^c_{aa}(t)$, right
after the measurement [see Fig.\ \ref{fig:trajectories}(a) and (b)],
based on a single-shot measurement result 
of $N_c(t)$ falling
into one of the two peak regions of $P(N,t)$  
[see Fig.\ \ref{fig:PNt}(a) and (b)]
even if the initial qubit state is unknown.
This perfect projective readout procedure, however, 
cannot be applied exactly to the case of 
non-vanishing $\Omega$. 
For $\Omega\neq 0$, there is always a chance for CQD qubit
not being exactly in the perfectly localized charge states after a 
single-shot measurement of time $t>t_{\rm m}$. 
This can be seen
from Fig.\ \ref{fig:trajectories}(c)-(f),
which show the appearance of wiggling or noisy portions
and the possibility of switching or making transitions in $\rho^c_{aa}(t)$
at some times $t>t_{\rm m}$
[i.e., $\rho^c_{aa}(t)$ does not stay and remain 
{\em exactly} at the value of $0$ or $1$].
The reason is that the occupation number operator of dot 1, 
$n_1=c_1^\dagger c_1$ (the measured quantity),
does not commute with the $\Omega$ term in
the Hamiltonian describing the free evolution of the qubit. 
Hence the free evolution, for $\Omega\neq 0$ case,
will interfere with the process of 
projection into the qubit charge states (logical 0 and 1 states) during the
readout measurement. 
Only when $\Omega=0$, $[n_1(t), n_1(t')]=0$ in the interaction picture.
This condition ensures that in this case,
if the qubit is projected into one of the
charge states [an eigenstate of $n_1(t_0)$], it remains in this state for all
subsequent times.
This is referred to as a quantum nondemolition measurement. \cite{Milburn94}

In reality, one may regard the single-shot readout to be possible 
either when the state-changing probability is lower than 
or the fidelity of the 
measurement is greater than their respective certain required
thresholds.
In the present CQD/PC model, the minimal condition to satisfy the 
requirement might be $t_{\rm mix}\gg t \gg t_{\rm m}$.
For $\Omega\neq 0$, we can switch on the energy mismatch  
${\cal E}\gg \Omega$ so that
the eigenstates of the qubit free-evolution Hamiltonian
are close to the qubit charge states. In this case,  
the probability of the qubit making transitions between the two 
charge states will be greatly reduced.
If the coupling between the detector and the qubit 
$\Gamma_d=|{\cal X}|^2/2$ can also be increased, the ratio of   
$(t_{\rm mix}/t_{\rm m})$ will increase further.
As a result, the fidelity of the single-shot readout
can be improved significantly.

Case II:
To be able to read out the 
initial qubit state, the bandwidth of the detector 
must be at least larger than the mixing rate, 
$\Delta \omega > t^{-1}_{\rm mix}$. 
As to read out the initial state of the CQD qubit in a single shot,
it is possible provided that   
(a) $(t_{\rm mix}/t_{\rm m})\to \infty$ and 
(b) $\rho_{aa}(0)=1$ or $\rho_{aa}(0)=0$, i.e.,  
the initial qubit state is in one of the
two perfectly localized states.
Note that condition (a) for symmetric CQD's (${\cal E}=0$) is equivalent to  
$(\Omega/\Gamma_d)\to 0$.
It is possible to distinguish the the two distributions
in $P(N,t)$ for $(\Omega/\Gamma_d)=0$ and $(\Omega/ \Gamma_d)=0.05$ 
cases in Fig.\ \ref{fig:PNt}(a)-(d). 
But there is still some overlap between the two distributions in the
$(\Omega/ \Gamma_d)=0.05$ case where $(t_{\rm mix}/t_{\rm m})\approx 25$. 
If the single-shot measurement result 
of $N_c(t)$ falls
into the overlap region, we will not be able to tell that the result
should contribute to the qubit being in dot $1$ or dot $2$
[see Fig.\ \ref{fig:PNt}(c) and (d)]. 
Although the probabilities for the ``ambiguous'' results to occur (or  
equivalently the state-changing or switching probabilities)
for the $(\Omega/ \Gamma_d)=0.05$ case, 
in Fig.\ \ref{fig:PNt}(c) with 
$t_{\rm mix}:t:t_{\rm m}\approx 25:10:1$
and in Fig.\ \ref{fig:PNt}(d) with 
$t_{\rm mix}:t:t_{\rm m}\approx 25:20:1$,
are rather small, they do not vanish since $\Omega\neq 0$.
Again, in practice, one may neglect these probabilities if they are
smaller than certain required probability thresholds.

However, even we can obtain a distinct result of $N_c(t)$
falling into one of the two
peak regions rather than the valley region 
[see Fig.\ \ref{fig:PNt}(a) and (b)]
for $\Omega=0$ case where $(t_{\rm mix}/t_{\rm m})\to \infty$. 
But if the initial qubit state is not in one of the
two perfectly localized states
[i.e., $\rho_{aa}(0)=1$ or $\rho_{aa}(0)=0$], 
we still do not know the initial state
based on the single-shot measurement result.
This is because in this case repeating the experiment 
many times to build up the probability distributions with weights 
corresponding closely to the initial values of the diagonal elements
of $\rho_{aa}(0)$ and $\rho_{bb}(0)$ is needed. 
We should also mention that even though we have determined 
$\rho_{aa}(0)$ and $\rho_{bb}(0)$
from the readout measurement, we still do not know if the
initial state is a mixed or pure state. We have to 
prepare the same initial state, and perform the readout
in different measurement basis, or  
perform single qubit rotations and then
repeat the experiment in the same charge basis of the 
measurement. \cite{White99} 
After this procedure, 
we may then be able to distinguish a pure initial state from a mixed one.

Case III:
Another situation is the measurement of a charge signal that undergoes
periodical classical probabilistic jumps, or random telegraph jumps,
between two different charge states. 
In this case the charge signal is in one of the two different charge states
at almost all times except at some very small times 
where the sudden switches of the charge state take place. 
As long as the charge signal 
would stay in one of the two charge states long enough
to provide enough signal-to-noise ratio ($t>t_{\rm m}$) and 
the largest frequency for charge state change 
(analogous to mixing rate $t^{-1}_{\rm mix}$)
is smaller than the bandwidth $\Delta\omega$, the
single-shot charge state readout for the charge signal 
may be possible. 
This charge signal could be generated by usual electronics device 
or by a charge defect
trap which could capture and emit an electron or other charged particles.
This case of classical charge signal is similar to the case 
of CQD charge qubit in the limit of vanishing $(\Omega/\Gamma_d)$ ratio
and with (initial) state in one of the two charge states. 
We can apply similar reasonings 
to the case of an incoherent 
electron charge transfer
between two quantum dots when an 
appropriate bias voltage is applied between them.
As long as the bandwidth
is larger than the sweeping bias frequency,
it is possible for this incoherent charge transfer 
to be observed in a single-shot measurement.

\newpage
\subsection{Possible experimental implementation}
\label{subsec:expt}
\noindent
Experimentally, coherent coupling between two CQD's has been reported.
It has been shown \cite{Crouch97,Waugh95}
that if the inter-dot tunneling barrier is low and the
strength of the coupling of two CQD's is strong, the two CQD's
behave as a large single dot in a Coulomb blockade phenomenon.
In addition, the
energy splitting between bonding and anti-bonding states of two CQD's
has been confirmed by microwave absorption measurements
\cite{Kouwenhoven98,Kouwenhoven00,Kouwenhoven01}.
The CQD qubit system discussed here may thus be fabricated as in
Ref.\ \cite{Kouwenhoven98,Kouwenhoven00}
and \cite{Kouwenhoven01} in a
AlGaAs/GaAs heterostructure system  
with external gates to control
the energy levels in each dot and tunneling coupling between the two
dots. We require strong inner- and inter-dot Coulomb
repulsion, so the double dot system can be tuned into a
regime \cite{Kouwenhoven98,Kouwenhoven00,Kouwenhoven01} 
such that effectively only one electron can occupy and tunnel
between the two dots and only one level in
each dot contributes to the dynamics.
Typical experimental values \cite{Kouwenhoven98,Kouwenhoven01}
for parameters $\Omega$ and ${\cal E}$
could be, for example, in the range of $0-200\mu {\rm eV}/\hbar$.        
It is also possible to fabricate the PC detector \cite{Buks98} in the same
heterostructure system. 
It was reported in a experiment\cite{Buks98}  with a
``which-path'' interferometer that Aharonov-Bohm interference is
suppressed owing to the measurement of which path an electron takes
through the double-path interferometer. A biased quantum point contact
(QPC) with arbitrary transparency
located close to a quantum dot, which is built in one of the
interferometer's arms, acts as a measurement device. The change of
transmission coefficient (probability) of the QPC, which depends on
the electron 
charge state of the quantum dot, can be detected.
The typical transmission probability $T_{\rm t}$ and 
average change in the transmission probability $\Delta T_{\rm t}$ 
for a low transparency PC at a source-drain voltage of 
$eV_{sd}=100\mu {\rm eV}$ could be \cite{Buks98}
in the range of $0.05-0.1$ and $0.013-0.028$, respectively.
The average tunneling rates $D$ and $D'$ is related to the
transmission probability as $D=|{\cal T}|^2=T_{\rm t}eV_{\rm sd}/\hbar$ and 
$D'=|{\cal T}+{\cal X}|^2=(T_{\rm t}-\Delta T_{\rm t})eV_{\rm sd}/\hbar$.
Note that the relative phase between ${\cal T}$ and ${\cal X}$ is 
$\theta=\pi$.

After a coherent manipulation, the qubit state is ready for readout.
We can then apply fast pulses\cite{Nakamura99}
on the external gates to tune, for example,  
$\Omega=1\mu {\rm eV}/\hbar$, ${\cal E}=20\mu {\rm eV}/\hbar$, 
and $eV_{\rm sd}=5 {\rm meV}$. If assuming $T_{\rm t}=0.05$ and 
$\Delta T_{\rm t}=0.028$, we find $D=250\mu {\rm eV}/\hbar$, 
$D'=110\mu {\rm eV}/\hbar$
and $t^{-1}_{\rm m}\approx \Gamma_d=(|{\cal X}|^2/2)\approx (14\mu {\rm
eV}/\hbar) \gg t^{-1}_{\rm mix}\approx 0.09\mu {\rm eV}/\hbar$.
It thus appears feasible to build a device to read out the initial
qubit state, as discussed in this paper.

However, the difficulty in realizing the quantum trajectories experimentally
is mainly due to the requirement of large bandwidth of the detector
signal coming out of the cryostat, i.e., $\Delta \omega \gg t^{-1}_{\rm
res}$.  We have not included the effect of other
environments (besides PC detectors) on the qubit dynamics during the
measurement process. It is assumed that the relaxation time induced by
these environments is much larger than the measurement time, $t_{\rm rel}\gg
t_{\rm m}$. But the environment-induced decoherence (dephasing) time may be
comparable to $(1/\Gamma_d)$, the detector-induced decoherence time in
the reduced density matrix approach. 
For example, the typical decoherence time for a 
single Cooper-pair-box charge qubit \cite{Nakamura99} is on the order of
$10-100$ ns, which implies that the bandwidth of the detectors should be 
at least of order of $1$GHz.

\section{Conclusions}
\label{sec:conclusion}
\noindent
We have discussed the connection between the 
quantum trajectory approach and master equation approach of the
reduced or ``partial'' reduced density matrix.
The quantum trajectory or stochastic
Schr\"odinger equation approach provides us with all the information  
as far as the system state evolution is concerned. 
The reduced or ``partially'' reduced density matrix 
can be obtained as a result of ensemble average or ``partial'' average
on the conditional, stochastic system density matrix 
over all possible detection records.
Each Quantum trajectory and corresponding detection record, 
mimics each possible single continuous in time 
measurement experiment.
Their possible individual realizations can
provide insight into, and aid in the
interpretation of, the ensemble average properties.
Especially, we have shown that the probability distribution 
$P(N,t)$ constructed from $10000$ realizations, is, as expected, in very good
agreement with that obtained from the Fourier analysis of 
the ``partially'' reduced density matrix.
In addition, the mixing behavior of $P(N,t)$ 
is illustrated and explained in
terms of different individual realizations of the
quantum trajectories and corresponding detection records. 
These results provide a unified picture 
for these seemingly different approaches reported in the literature.

We have also discussed
the possibility of a single-shot readout of the qubit state. 
In the present CQD/PC model, for ${\cal E}=0$ it may be possible in the
limit of vanishing $\Omega/\Gamma_d$ ratio.
Generally speaking, in order to obtain a confident CQD qubit state readout
in the charge state basis, it is 
better to (a) reduce the coherent coupling $\Omega$ between the charge state,  
(b) increase the interaction $\Gamma_d$ with the PC detector
and (c) switch on the energy mismatch ${\cal E}$ 
for the readout measurement.

\nonumsection{Acknowledgments}
\noindent
The author would like to thank G.~J.~Milburn for his
suggestions and comments on the manuscript. 
Useful conversations with S.~A.~Gurvitz, G.~Sch\"on, A.~G.~White
and H.~M.~Wiseman are acknowledged.
The author would also like to 
acknowledge support through the Hewlett-Packard Fellowship.

\nonumsection{References}
\noindent



\end{document}